\documentclass[aps,prd,twocolumn,superscriptaddress]{revtex4-2}
\usepackage{xspace}
\usepackage[colorlinks]{hyperref}

\usepackage{float}
\usepackage{graphicx}
\usepackage{epstopdf}
\usepackage{setspace}
\usepackage{bbold}
\usepackage{comment}
\usepackage{color}
\usepackage{soul}
\usepackage[abs]{overpic}
\usepackage{amsmath}
\usepackage{lipsum}
\usepackage[abs]{overpic}
\usepackage{nicefrac}
\begin{document}

\author{Jiaxiang Wang}
\affiliation{Wright Laboratory, Department of Physics, Yale University, New Haven, Connecticut 06520, USA}

\author{T. W. Penny}
\affiliation{Wright Laboratory, Department of Physics, Yale University, New Haven, Connecticut 06520, USA}

\author{Juan Recoaro}
\affiliation{Wright Laboratory, Department of Physics, Yale University, New Haven, Connecticut 06520, USA}

\author{Benjamin Siegel}
\affiliation{Wright Laboratory, Department of Physics, Yale University, New Haven, Connecticut 06520, USA}

\author{Yu-Han Tseng}
\affiliation{Wright Laboratory, Department of Physics, Yale University, New Haven, Connecticut 06520, USA}

\author{David C. Moore}
\affiliation{Wright Laboratory, Department of Physics, Yale University, New Haven, Connecticut 06520, USA}
\affiliation{Yale Quantum Institute, Yale University, New Haven, Connecticut 06520, USA}

\title{Mechanical detection of nuclear decays}

\begin{abstract}
We report the detection of individual nuclear $\alpha$ decays through the mechanical recoil of the entire micron-sized particle in which the decaying nuclei are embedded. Momentum conservation ensures that such measurements are sensitive to any particles emitted in the decay, including neutral particles that may otherwise evade detection with existing techniques. Detection of the minuscule recoil of an object more than $10^{12}$ times more massive than the emitted particles is made possible by recently developed techniques in levitated optomechanics, which enable high-precision optical control and measurement of the mechanical motion of optically trapped particles. Observation of a change in the net charge of the particle coincident with the recoil allows decays to be identified with background levels at the micro-Becquerel level. The techniques developed here may find use in fields ranging from nuclear forensics to dark matter and neutrino physics.
\end{abstract}

\maketitle

\paragraph*{Introduction.} 
Accurate detection of nuclear decays is central to a variety of fields in physics, engineering, and medicine. The vast majority of detectors for such decays rely on measuring the energy deposited by decay products, typically through generation of ionization, scintillation, or phonons (heat) in a detection medium~\cite{knoll2010radiation}.
However, these technologies ultimately rely on the decay products themselves to interact in the detector and deposit most of their energy, allowing particles that exit the detector to evade detection. A well-known example is nuclear $\beta$ decay, where the emitted neutrino will escape any practical detector, and only a fraction of the total decay energy is detected with conventional techniques. 

An alternative approach is to reconstruct the momentum imparted to the object containing the decaying nucleus by the recoil of its daughter, following the escape of the primary decay products~\cite{PhysRevA.98.052103}. 
This momentum-based reconstruction is sensitive to any escaping particles, including neutral particles, and has been demonstrated for individual recoiling nuclei~\cite{PhysRevC.58.2512, PhysRevLett.90.012501_trap}. However, the tiny recoil has been previously undetectable for a much more massive solid object containing the decaying nucleus. For example, the $\alpha$ decays detected here impart a momentum impulse of only $\sim 10^{-19}$~kg m/s to the micron-sized spheres containing the decaying nuclei. 

Detection of such tiny recoils requires extreme isolation of the object containing the decaying nucleus from thermal noise and precise measurement of the object's motion. These stringent requirements can now be met by rapid progress in the optomechanical control and measurement of levitated particles~\cite{Levitodynamics2021}. By optically trapping nano- to micron-size particles in high vacuum, thermal noise can be made negligible~\cite{2020Millen_review}. 
The interaction of the trapped particle with the laser used to detect its position then provides the dominant noise source~\cite{Monteiro:2020qiz_force, Levitodynamics2021, 2020Millen_review}. Ultimately, the minimal possible measurement induced noise is constrained by quantum mechanics~\cite{Clerk_review, Beckey_review}, and particles as large as a femtogram in mass (100~nm in diameter) are now reaching the quantum measurement regime~\cite{Delic2020, magrini2021real, Tebbenjohanns2021, Ranfagni:2021pnb, Kamba:2022cgy, Piotrowski:2022qda}. Such particles are extremely precise force sensors~\cite{liang2023yoctonewton,Moore_2021}, finding applications in tests of quantum mechanics with massive objects~\cite{PhysRevLett.107.020405, Roda-Llordes:2023odc, PhysRevLett.127.023601, Neumeier:2022czd_mario, 2018NJPh...20l2001S}, searches for new short-distance interactions~\cite{PhysRevD.104.L061101_gratta_grav, PhysRevLett.105.101101}, dark matter~\cite{Riedel2013, CarneyMechanical, PhysRevLett.125.181102, Afek:2021vjy_coherent}, gravitational waves~\cite{aggarwal2020searching, PhysRevLett.129.053604}, and neutrino physics~\cite{PRXQuantum.4.010315}. 

Here we demonstrate the detection of individual nuclear $\alpha$ decays using optically levitated micron-sized spheres (with mass $>10$~pg). Decays are detected through the change in the net electric charge of the particle following the decay and the coincident recoil of the entire particle. The combination of these signatures allows both extremely low backgrounds and sensitivity to the momentum carried by the decay products through the measurement of the recoil. While these techniques already reach signal-to-noise ratios above 10 for the recoil measurement, more than $100 \times$ further improvement in the momentum sensitivity is expected as these larger spheres are also brought into the quantum measurement regime. 

\begin{figure*}
    \centering
    \includegraphics[width=\textwidth]{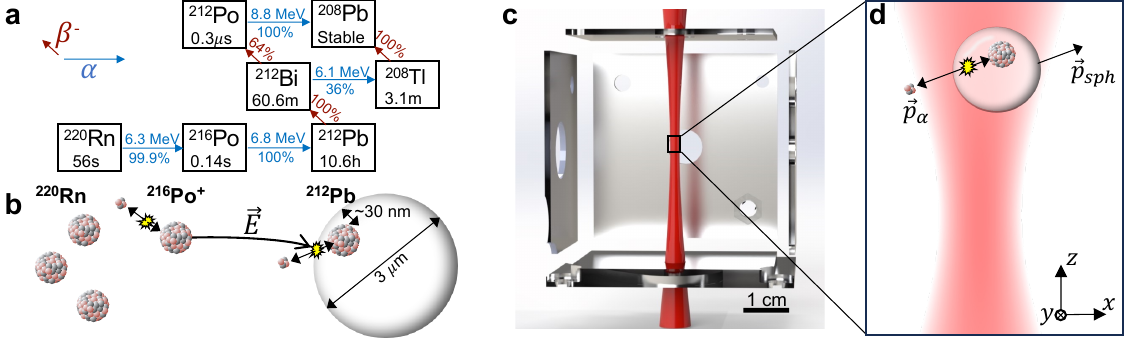}
    \caption{(a) $^{220}$Rn decay chain~\cite{nudat}. For each isotope, the decay half-life is given within the box. Arrows indicate the primary $\alpha$ decay energies and branching ratios (blue), as well as the relevant $\beta$ decay branching ratios (red). (b) Schematic of $^{212}$Pb implantation procedure described in the text. (c) Cross-sectional view of the electrodes surrounding the trap (an additional electrode on the front face and two weak imaging beams are not pictured). (d) Schematic of a sphere recoil following an $\alpha$ decay of $^{212}$Bi or $^{212}$Po, for which the daughter nucleus is stopped within the sphere while the $\alpha$ (and additional low-energy $e^-$ and possibly secondary nuclear recoils, not pictured) exit the sphere. In this case, the momentum of the $\alpha$ particle can be inferred from the sphere recoil ($\vec{p}_{\alpha} = -\vec{p}_{sph}$).}
    \label{fig:setup}
\end{figure*}

\paragraph*{Experimental description.} In our experiment, we detect single nuclear decays occurring in silica spheres of radius $\approx 1.5\ \mu$m that are optically trapped in vacuum at pressures between $10^{-8}$ to $10^{-7}$~mbar. Before trapping, the spheres are implanted with $^{212}$Pb, an unstable radon daughter in the thorium decay chain (Fig.~\ref{fig:setup}~a). $^{220}$Rn is introduced into an implantation chamber~\cite{supplement} where it undergoes $\alpha$ decay to produce $^{216}$Po$^+$ ions, which are drifted by an electric field to the sphere surfaces (Fig.~\ref{fig:setup}~b). The $^{216}$Po $\alpha$ decays then implant $^{212}$Pb daughters into the spheres at depths $\lesssim 60$~nm. The implanted spheres are then transferred to a high-vacuum chamber and loaded into an optical trap formed by a focused laser, where the subsequent decays are measured.

The optical trap is surrounded by six planar electrodes (Fig.~\ref{fig:setup}~c). The lower electrode is grounded, while the upper five can be biased independently. The electrodes and trapping beam define a coordinate system with the trapping beam propagating vertically in the $z$ direction, and the $x$ and $y$ directions defined to be perpendicular to the electrode faces (Fig.~\ref{fig:setup}~d). Once trapped in high vacuum, the sphere's net electric charge is determined by monitoring the sphere's motion in response to an oscillating electric field applied using these electrodes~\cite{Monteiro:2020qiz_force}. The net charge is calibrated into units of a single elementary charge, $e$, by observing discrete steps in the response~\cite{Monteiro:2020qiz_force, Frimmer2017, Moore2014}, with the phase of the motion relative to the electric field determining the polarity. Electrons can be added to the sphere via thermionic emission from a tungsten filament or removed from the sphere via an ultraviolet (UV) lamp. The trapping beam is circularly polarized to rotate the sphere to $\gtrsim 100$~kHz rotational frequencies, gyroscopically stabilizing its motion and eliminating noise from fluctuations in its angular orientation~\cite{Monteiro:2020qiz_force, Monteiro2018Rotation}. The electrodes contain holes for the trapping beam and two weakly focused perpendicular imaging beams. The sphere position is detected by imaging the transmitted light on three detectors (one for each of $x$, $y$, and $z$). The dominant noise source in the $x$ direction at the base pressure of $2\times 10^{-8}$~mbar is thermal noise from residual gas. The $y$ and $z$ directions suffer from additional technical noise due to fluctuations in the trapping beam pointing and intensity~\cite{Monteiro:2020qiz_force}.

\begin{figure*}
    \centering
    \includegraphics[width=\textwidth]{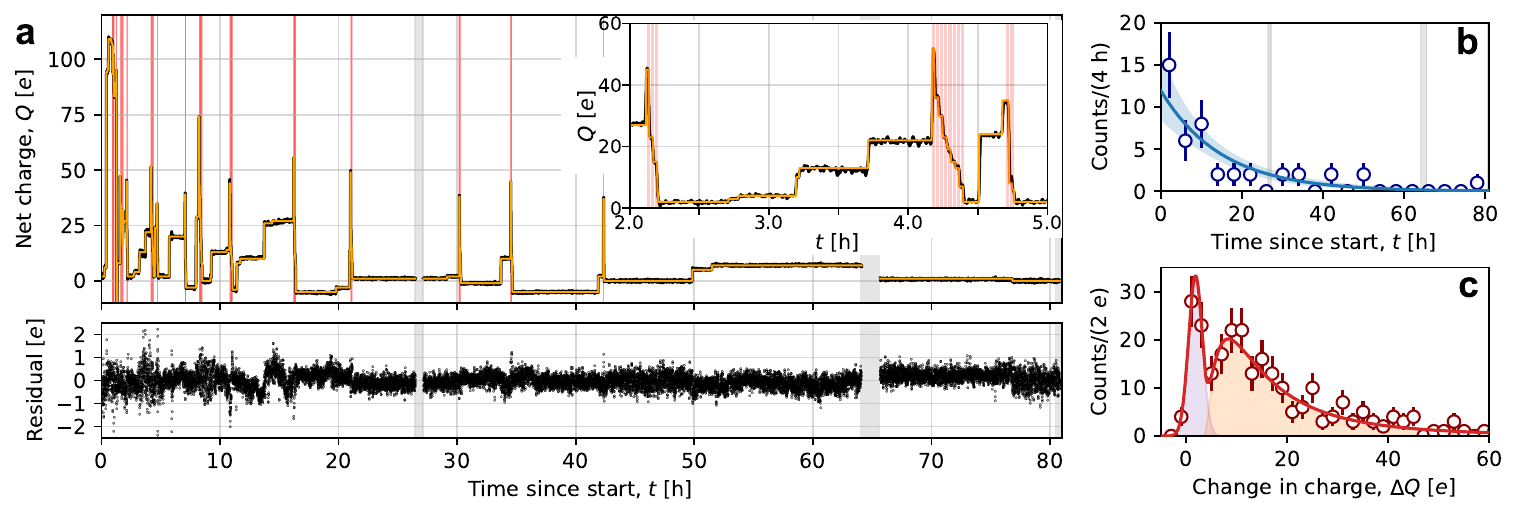}
    \caption{(a, upper panel) Example of the measured charge versus time for a sphere implanted with $^{212}$Pb. Positive values of the charge correspond to a net excess of protons over electrons in the sphere. The black line shows the measured sphere charge (averaged in 12~s intervals) while the orange line shows the best fit to these data after reconstructing the location of each charge change. Light red vertical lines show when the filament adding $e^-$ to the sphere was active, while vertical gray bands indicate dead time during calibrations. The inset shows a zoom to several charge changes near the beginning of the data. (a, lower panel) Residual between the best-fit charge and the measured data. The residuals are typically $<1\ e$, demonstrating sensitivity to changes in the net charge of the sphere by a single $e$. (b) Example fit to the reconstructed number of charge changes versus time for the data shown on the left (blue points, with statistical error bars), with best fit $T_{1/2} = 10.3^{+1.8}_{-1.5}$~h (blue band). (c) Distribution of reconstructed charge changes for all spheres studied in this work, which is empirically consistent with a sum (red line) of the two components (shaded) described in the text. }
    \label{fig:charge}
\end{figure*}

\paragraph*{Charge measurement.} 
The results presented here consist of data from six spheres implanted with $^{212}$Pb following the procedure above. Each sphere is continuously measured for 2--3 days after loading into the optical trap and pumping to high vacuum. Once in vacuum, the charge is continuously monitored and individual decays occurring within the sphere are identified by detecting a change in the net charge of the sphere, $\Delta Q$. Figure~\ref{fig:charge}~a shows an example of the net charge of the sphere over time after reaching $<10^{-7}$~mbar.

An automatic discharging procedure is implemented to maintain the net charge $|Q| < 50\ e$. Discharging periods are recorded by the data acquisition (DAQ) system and excluded from the charge analysis (red lines in Fig.~\ref{fig:charge}~a). Dead time due to the DAQ, impulse response calibrations, and these discharging periods is $<$10\% of the total measurement time for all spheres considered.

For a typical implanted sphere, the decay frequency observed from these charge changes is initially around 2--5 decays per hour, with a decrease consistent with the $^{212}$Pb half-life, $T_{1/2} = 10.6$~h (see Fig.~\ref{fig:charge}~b). The background rate of charge changes was measured for an unimplanted sphere, with no charge changes observed in 3~days. As a result, any charge changes can be used to identify decays with backgrounds at the $<1/(\mathrm{day}) \sim \mu$Bq level. For the measurements presented here, this charge measurement is crucial to reject transient noise bursts~\cite{supplement}, which were found in previous work to be associated with vibrational or acoustic noise~\cite{PhysRevLett.125.181102}.

The distribution of charge changes from data across all spheres is shown in Fig.~\ref{fig:charge}~c. While a detailed study of this distribution is beyond the scope of this work, several qualitative features are apparent. First, almost all charge changes are measured to be in the positive direction---i.e., the spheres nearly always lose more electrons than protons for both the $\alpha$ and $\beta$ decays in the decay chain. Second, two peaks in the charge distribution are observed, and the data are empirically consistent with a Gaussian-like component peaking around $\Delta Q = +2$~$e$ and a heavy-tailed component (here modeled by a log-normal distribution) peaking around $\Delta Q = +8$~$e$, but with a tail of events extending to $\Delta Q \gg +50$~e.
Of the 257 total observed decays, 9 were observed with charge changes $>60\ e$ (beyond the range plotted in Fig.~\ref{fig:charge}~c), with the largest observed change of $\Delta Q = +148\ e$.

\paragraph*{Recoil measurement.} The observed charge changes are used to tag decay events and provide timing information to search for recoils from the momentum transferred to the sphere during a decay. Since the time for all decay products to exit or stop within the sphere is much faster than its $\sim$10 ms mechanical response time, the decay will impart an essentially instantaneous momentum impulse, which can be detected through the subsequent motion of the sphere's center-of-mass.

To reconstruct the response of the sphere to such impulses, {\em in situ} calibrations are performed using electric impulses of known amplitude. Using the known net charge of the sphere and finite element method simulations of the electrode geometry, this calibration allows the impulse response to be measured to a relative accuracy of $<2$\% in the $x$ and $y$ directions, and to $<10$\% accuracy in the $z$ direction, limited by uncertainty in the knowledge of the sphere position relative to the electrodes~\cite{supplement}. 

\begin{figure}[t]
    \centering
    \includegraphics[width=\columnwidth]{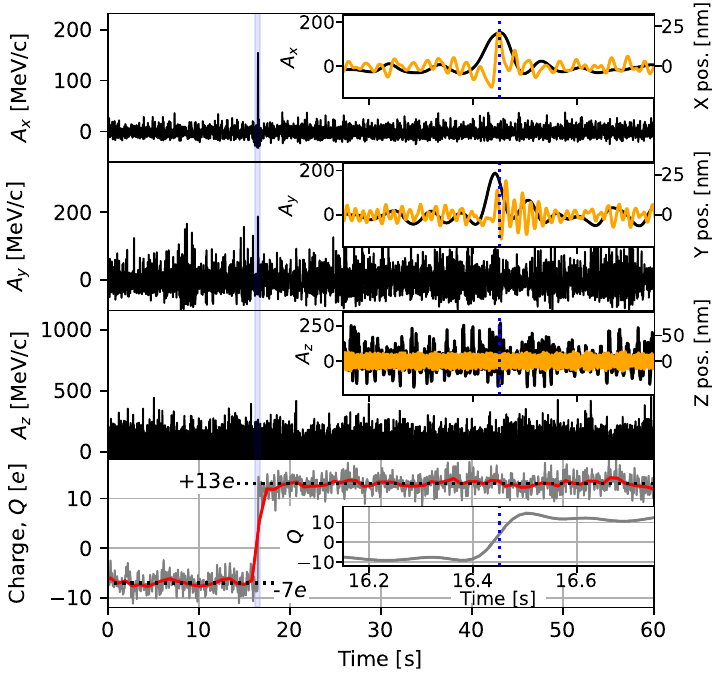}
    \caption{Example reconstructed sphere recoil coincident with a charge change. The upper three panels show the filtered magnitude of the reconstructed impulse amplitude, $A_j$, for $j = x, y, z$ versus time. The insets show a zoom to the region within $\pm 0.3$~s of the reconstructed charge change time (light blue region in outer panel). A recoil is reconstructed in the $x$ and $y$ directions with a combined signal-to-noise ratio of 13 at the time indicated by the blue dotted line. The orange curve shows the reconstructed position (right axis) versus time, filtered around the resonant frequency for each direction, showing the expected damped harmonic oscillator response for $x$ and $y$ (and $z$ response consistent with noise). The lower panel shows the reconstructed charge averaged over 25 ms (gray) and 1.6 s (red) windows.}
    \label{fig:pulse}
\end{figure}

An ``optimal filter''~\cite{Gatti:1986cw,supplement}, making use of the calibrated impulse response and measured noise spectrum for each sphere, was found to provide the best resolution for impulse amplitude reconstruction. In the $x$ direction, a resolution in the range of $\sigma_x = 15 - 27$~MeV/c was measured for the spheres considered. This corresponds to an expected signal-to-noise ratio between $9 - 17$ for a $^{212}$Po $\alpha$ decay with energy $E_\alpha = 8.8$~MeV imparting its momentum $\sqrt{2 m_\alpha E_\alpha} = 256$~MeV/c along the $x$ direction, where $m_\alpha$ is the $\alpha$ particle mass. The resolution measured in the $y$ and $z$ directions is found to be poorer than in the $x$ direction in our current system due to the larger impact of technical noise.
 
Figure~\ref{fig:pulse} shows a candidate $\alpha$ decay coincident with an observed change in the charge of the sphere. The impulse amplitude reconstructed from the optimal filter, and calibrated using the {\em in situ} calibration described above, shows a recoil of the sphere coincident in time with the charge change (whose time can be reconstructed with $<100$~ms precision), consistent with the impulse that would be expected for an $\alpha$ decay aligned between the $x$ and $y$ directions.

\begin{figure}[t!]
    \centering
    \includegraphics[width=\columnwidth]{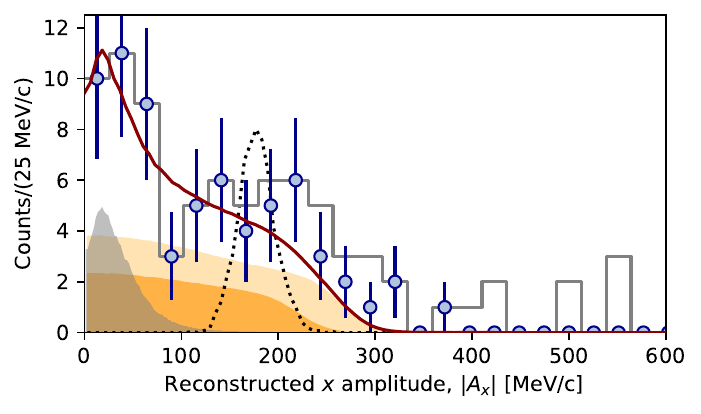}
    \caption{Spectrum of the reconstructed $x$ impulse amplitudes for each observed charge change. All data are shown by the gray histogram, while events with $\Delta Q < 30\ e$ are shown as blue points with error bars. The best fit to the spectrum (dark red line) is a sum of the predicted response from $^{212}$Bi $\alpha$ decays (dark orange shaded), $^{212}$Po $\alpha$ decays (light orange shaded), and $\beta$ decays that impart momentum consistent with random sampling of the noise (gray shaded). The black dotted line shows the expected resolution for impulses in the $x$ direction measured from the {\em in situ} calibrations. }
    \label{fig:spectrum}
\end{figure}

The distribution of reconstructed impulse amplitudes in the $x$ direction coincident with detected charge changes for all spheres considered here is shown in Fig.~\ref{fig:spectrum}. Compared to the charge data summarized in Fig.~\ref{fig:charge}~c, this distribution contains fewer events (83 out of 257 total events) since only data periods at sufficiently low pressure that the sphere had reached a rotational velocity $>100$~kHz and the noise was stable were considered for the recoil analysis, while all time periods were used in the charge analysis. Two of the six spheres were also not considered in this recoil analysis since they lacked {\em in situ} impulse calibrations. Only the $x$ projection of the momentum is considered in the recoil analysis due to its higher signal-to-noise (and observed systematic drifts in the $y$ amplitude between calibrations).

In Fig.~\ref{fig:spectrum}, all reconstructed recoils are shown by the gray line, while only recoils coincident with charge changes for which $\Delta Q < 30\ e$ are indicated by the blue points. The reconstructed amplitudes are corrected for the small impulse induced by the charge monitoring field in the $x$ direction, which is estimated to provide negligible error after correction. The data with $\Delta Q < 30\ e$ are consistent with the expected distribution of momenta projected in the $x$ direction based on a Monte Carlo simulation of the $\alpha$ decays from $^{212}$Bi and $^{212}$Po~\cite{supplement}. In addition to these $\alpha$ decays, the decay chain contains several $\beta$ and $\gamma$ transitions, which may occur in coincidence with the $\alpha$ decay. While these particles can carry non-negligible energy compared to the $\alpha$, their relative momentum is much lower ($\sim$1--3~MeV/c), and is not detectable with the resolution of the existing setup. The maximum momentum transfer occurs when the $\alpha$ particle escapes along the $x$ direction and the daughter nucleus stops within the sphere. 

A small fraction of the observed decays extend beyond the maximum momentum expected from these simulations. The cut removing events with $\Delta Q < 30\ e$ in Fig.~\ref{fig:spectrum} indicates that only events in the tail of the charge change distribution contribute to these unexpectedly large decays. Systematic checks were performed to ensure these large reconstructed recoils were not due to miscalibration or reconstruction errors.
While further study is required to definitively identify the origin of these recoils, these events may arise from decays in which the nuclear recoil is emitted along the sphere's surface, leading to a large number of secondary nuclear recoils ejecting Si and O atoms. These ejected atoms may carry some fraction of their momentum in the direction of the emitted $\alpha$, and might also produce the observed large changes in the net charge of the sphere as secondary electrons escape.

\paragraph*{Discussion and outlook.} This work demonstrates the detection of single nuclear decays in optically trapped, micron-sized spheres through both the change in the sphere's charge and its coincident recoil. While this initial work focuses on $\alpha$ decays, these techniques become especially powerful for decays in which weakly interacting neutral particles would escape conventional detectors. Extending the same techniques to femtogram mass spheres will allow reconstruction of the momentum of a single neutrino leaving the sphere with signal-to-noise $>$100~\cite{PRXQuantum.4.010315}. Such techniques are generically sensitive to any ``invisible'' massive particles emitted in nuclear decays, including sterile neutrinos~\cite{PRXQuantum.4.010315} or particles that may be related to dark matter~\cite{Benato:2018ijc, Dent:2021jnf}.

Beyond fundamental physics, these techniques may find applications in nuclear forensics, which aims to determine the isotopic composition of a nuclear material~\cite{PhysRevA.98.052103, KRISTO2020921}. While future work is required to fully characterize what background levels may be possible, from the measurements presented here it appears plausible that sub-$\mu$Bq background rates will be achievable, possibly permitting detection of long-lived species such as $^{235}$U in single captured dust particles~\cite{PhysRevA.98.052103}. 
The measurements presented here will also allow the characterization of ejected radon daughters, as well as low-energy secondary electrons and ions, from decays near a solid surface, which may be relevant for applications in nuclear medicine~\cite{majkowska2020nanoparticles, Holzwarth_alpha_escape} and rare event searches (e.g.,~\cite{Wandkowsky:2013una, LZ:2022ysc, Meng_2022}). Finally, the radon daughter decays ejecting $>100$ low-energy $e^-$ from a surface observed here may ultimately be observable in ion-based quantum computers, where a large number of charged particles passing through an ion array could simultaneously interact with multiple qubits~\cite{Carney2021Trapped}. Similar to phonon-mediated particle interactions in superconducting qubit arrays~\cite{McEwen:2021wdg}, these events may become evident only as error correction techniques begin to suppress uncorrelated errors and these systems are scaled to large size.

While this initial work has already demonstrated detection of single nuclear decays with signal-to-noise $>10$, substantial further improvement is expected. Future work is required to ensure the signal-to-noise is independent of the decay direction, characterize the background rates achievable, integrate conventional particle detectors around the trap, and to reach---and eventually surpass~\cite{PRXQuantum.4.030331_squeeze,PhysRevX.13.041021_ligo}---the ``standard quantum limit'' for the detection of the sphere recoil~\cite{PhysRevB.70.245306, Beckey_review}, where the momentum resolution would be $\sqrt{\hbar m \omega_0} \approx 50$~keV/c for the spheres considered here (with mass $m \sim 10$~pg and angular resonant frequency $\omega_0 \sim 2\pi \times 100$~Hz). The ongoing rapid progress in the field of levitated optomechanics promises to extend the future sensitivity of these techniques by orders-of-magnitude. 

\paragraph*{Acknowledgments.} We thank Daniel Carney and Giorgio Gratta for useful discussions, and Fernando Monteiro and Gadi Afek for early work on the techniques used here. This work was supported through the DOE Office of Nuclear Physics, Quantum Horizons Award DE-SC0023672, and in part by ONR Grant N00014-23-1-2600 and NSF Grant PHY-2109329.

\bibliography{AlphaRecoils}

\clearpage
\onecolumngrid
\section*{Supplemental Material}\label{sec:suppl}

\subsection{Optical setup}
\label{sec:setup}
\begin{figure*}[b]
    \centering
    \includegraphics[width=1.0 \textwidth]{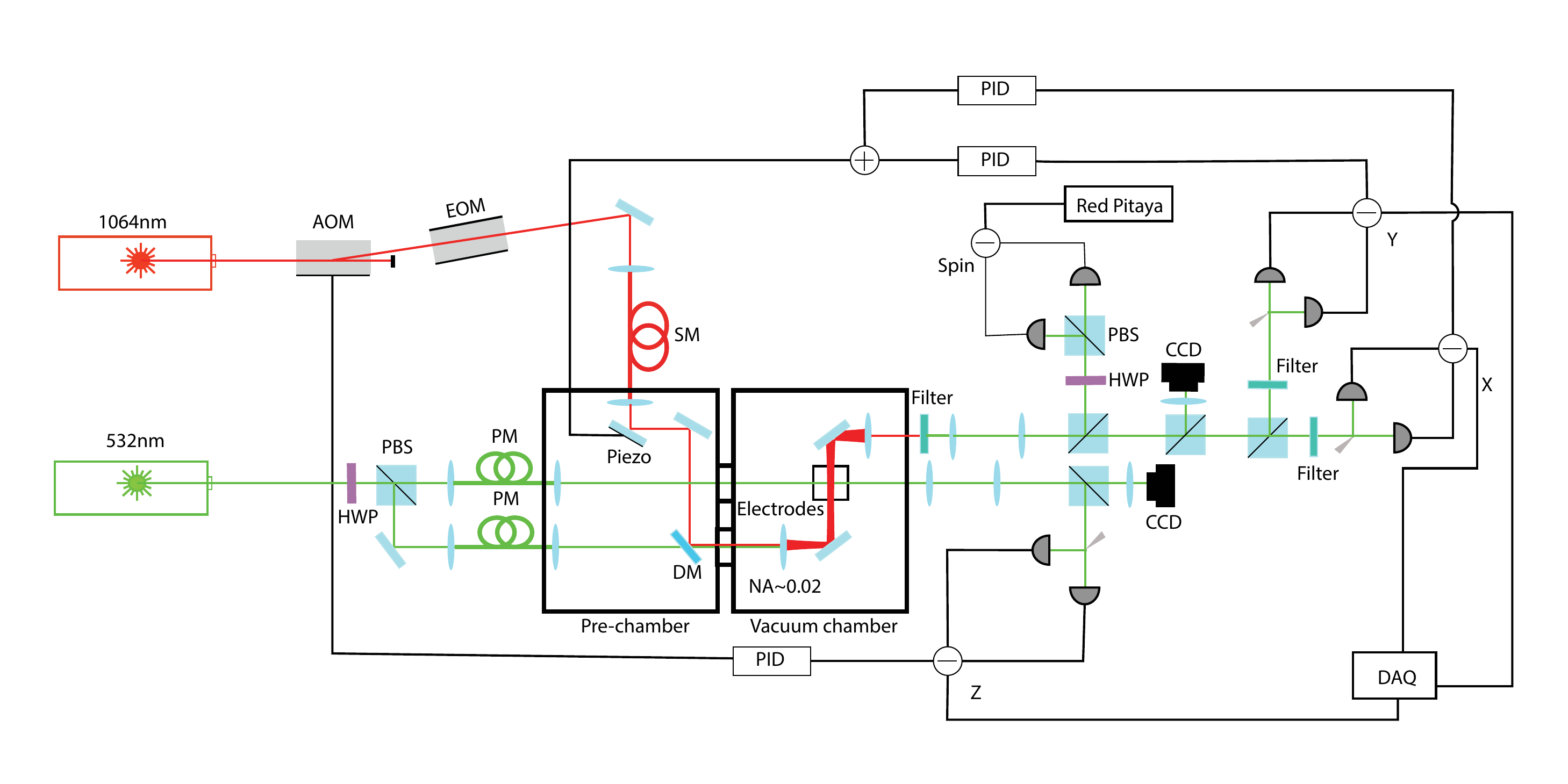}
    \caption{Optical diagram of the experimental setup described in the text}
    \label{fig:optics}
\end{figure*}

The optical trap is formed by a 1064 nm vertically oriented trapping beam. The beam's output power is modulated by an acousto-optic modulator (AOM), and its polarization is altered from linear to circular using an electro-optic modulator (EOM). The beam is then coupled to a single-mode  (SM) fiber leading to a sealed pre-chamber. Here, a 2-axis piezo-controlled mirror (Thorlabs ASM003) adjusts the beam's direction. The trapping beam, with an approximate power of 15 mW, then enters the trapping chamber. It is vertically oriented, and weakly focused with a numerical aperture (NA) $\approx 0.02$ to form the trap at the center of the electrodes. The 1064~nm beam is immediately blocked after leaving the vacuum chamber by a shortpass filter and cleaned by notch filters.

For detection, two linearly polarized 532 nm beams, split from a single laser source by a polarized beam splitter (PBS), are used. After coupling to polarization-maintaining (PM) fibers, these beams enter the pre-chamber and finally the trapping chamber. One beam is coaligned with the trapping beam via a dichroic mirror (DM) and directed vertically onto the sphere to collect information about $x$, $y$ displacement and rotation of the sphere. The second beam is oriented horizontally and gathers information about the $z$ displacement~\cite{Monteiro:2020qiz_force}. Three D-shaped mirrors split the sphere's image along three axes for data collection, with balanced photodiodes measuring the relative intensity in each half of the image. A data acquisition system (DAQ) records the voltage signals from the balanced diodes with a sampling rate of 10~kHz. A half-wave plate (HWP) and a PBS followed by a balanced photodiode are used to determine the rotational orientation of the sphere~\cite{Monteiro2018Rotation}, with the balanced photodiode voltage digitized at 125~MHz by a Red Pitaya.

A field-programmable gate array (FPGA) is used to implement proportional–integral–derivative (PID) feedback, based on signals from the photodiodes, to control the sphere's center-of-mass motion in three degrees of freedom. Modulation of the AOM drive regulates the trapping beam's power to provide feedback in the $z$ direction. Feedback in the $x$ and $y$ directions is applied via the piezo-controlled mirror, which displaces the trapping laser relative to its equilibrium position. For this work, the feedback in the $x$ direction is implemented primarily using the derivative term, providing damping but no change to the resonant frequency. In the $y$ and $z$ directions, in addition to damping through the derivative term, there is a significant proportional gain applied by the feedback that leads to stiffening of the optical spring constant, which separates the resonant frequencies in these directions from the $x$ resonant frequency. For some spheres, a non-zero integral gain is applied in the feedback loop to stabilize long-term drifts over the multi-day measurements described here.

\subsection{$^{212}$Pb loading}
\label{sec:pb_loading}
\begin{figure*}
    \centering
    \includegraphics[width=1.0 \textwidth]{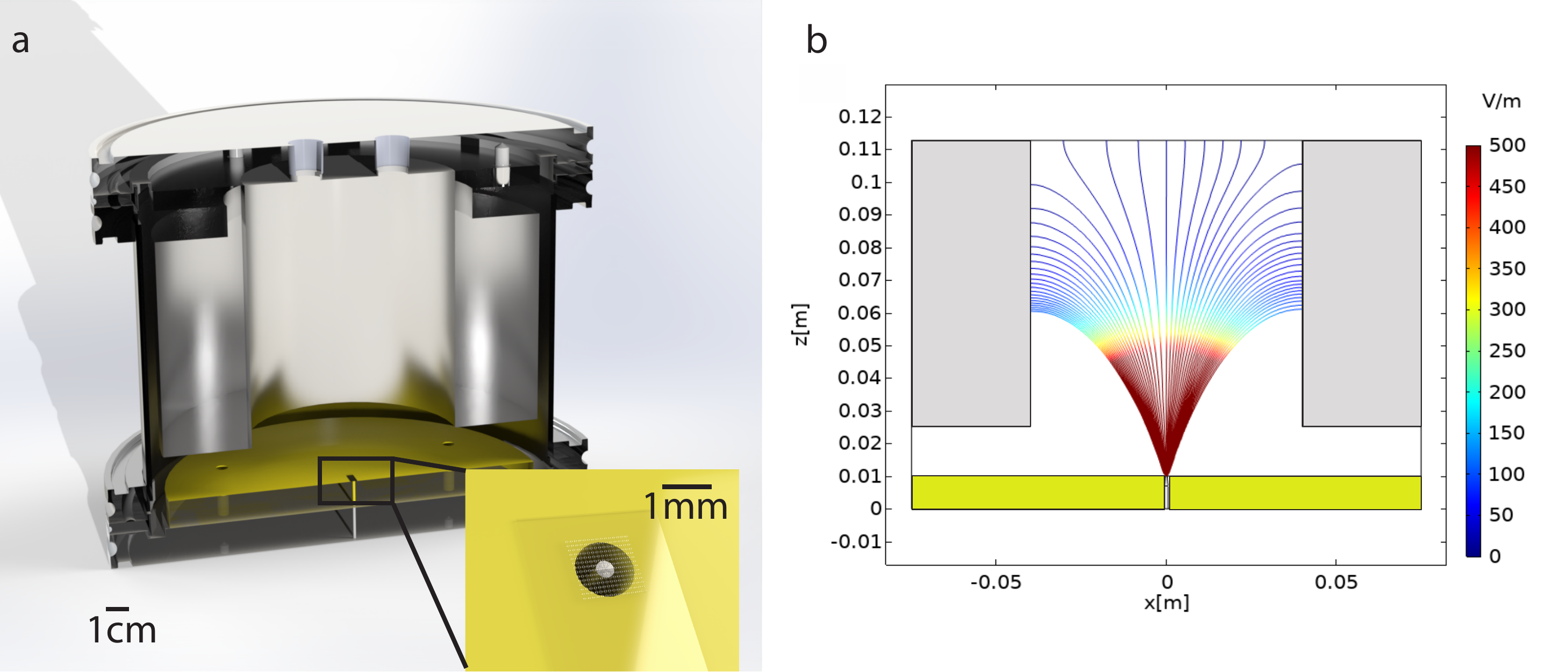}
    \caption{(a, left) Schematic of the implantation chamber. A metal pin is placed at the bottom center of the chamber to apply the DC high voltage. A Delrin disk surrounds the pin, securing the sphere-loaded glass cover slip and stabilizing the pin's position. Two ports on the chamber's top lid allow gas inflow and outflow. The inset provides a magnified view of the spheres on the upper surface of the cover slip, with the pin positioned underneath. (b, right) Cross-section of a simulation depicting the electric field lines within the implantation chamber, illustrating the percentage of the volume directed to the approximately 1~mm$^2$ region containing the spheres when the pin is biased to a voltage of $-1$~kV. The gray region is an aluminum spacer, which, along with the chamber walls, is grounded. The yellow region is the Delrin disk, which provides mechanical support and electrical insulation of the pin.}
    \label{fig:loading_chamber}
\end{figure*}

Commercially produced silica microspheres are used in this work (Corpuscular, C-SIO-3.0). The mean radius of the spheres measured by an optical microscope is $1.4$~$\rm{\mu}$m, with a variation of 5\% between different spheres. The spheres are loaded with $^{212}$Pb using a flow-through $^{220}$Rn emanation source (Pylon TH-1025, activity of 10 kBq on Jan. 1, 2023). Dry microspheres are applied to the surface of a glass cover slip, which is positioned above a negatively biased electrode in a dedicated vacuum chamber for the implantation process. The chamber is pumped to a pressure of 10 mbar and maintained at this pressure while $^{220}$Rn is transported to the chamber by a flow of 4.6 sccm of nitrogen carrier gas. Simulations using the Stopping and Range of Ions in Matter (SRIM) software package~\cite{ziegler2008srim, srim} indicate this pressure is sufficient to stop the majority of $^{216}$Po recoils following decay of the $^{220}$Rn nucleus before they hit the chamber walls. Previous measurements indicate a substantial fraction of $^{216}$Po daughters are expected to remain as singly ionized $^{216}$Po$^+$ after stopping in dilute gas~\cite{hopke1996initial, PAGELKOPF20031057}. These ions are drifted to the surface of the cover slip on which the spheres are deposited by the electric field from a 250~$\mu$m diameter pin positioned just below the opposite side of the cover slip and biased to a potential of $-1$~kV (see Fig.~\ref{fig:loading_chamber}). 

\begin{figure*}
    \centering
    \includegraphics[width=0.8\textwidth]{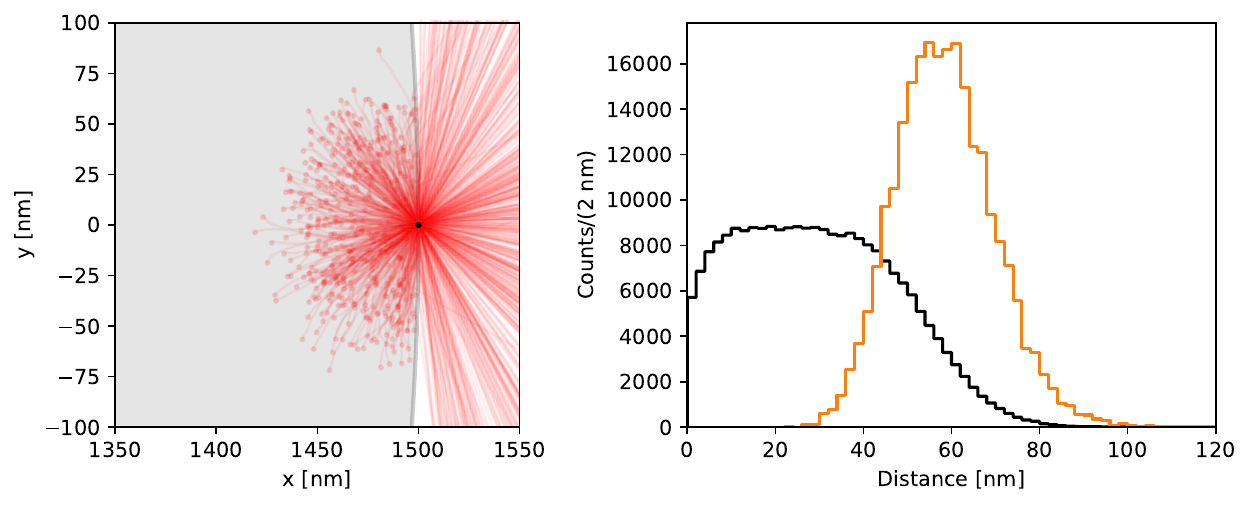}
    \caption{(left) Simulation of the implantation of $^{212}$Pb daughters into the microspheres following $\alpha$ decays of $^{216}$Po on the sphere surface. The red lines indicate the projection into the $x$-$y$ plane of simulated trajectories of randomly oriented $^{212}$Pb daughter recoils starting from $(x,y,z) = (1.5, 0, 0)\ \mu$m. Approximately 50\% of decays are stopped within the sphere (gray). (right) Distribution of implantation depths as a function of the nearest distance from the sphere surface (black). The median simulated implantation depth is 29~nm, with 95\% of $^{212}$Pb implanted within 60~nm of the surface. The orange distribution indicates the total distance between the $^{216}$Po decay location and final $^{212}$Pb location.}
    \label{fig:implant_depth}
\end{figure*}

The spheres are loaded with $^{220}$Rn daughters for a total period several times longer than the $^{212}$Pb half-life, so that secular equilibrium is reached and the maximum sphere activity is achieved (typically 1-2 days). The typical measured sphere activity following loading translates to a average activity of 200~Bq/mm$^2$ of $^{216}$Po$^+$ reaching the surface of the cover slip (translating to approximately 1 mBq activity per sphere). The $^{216}$Po$^+$ ions drifted to the sphere surface promptly decay and, as shown in Fig.~\ref{fig:implant_depth}, approximately 50\% of the time these decays implant the $^{212}$Pb daughter into the surface of the sphere at a depth up to 60~nm. 

After loading, the cover slip is positioned in the main trapping chamber above the optical trap and spheres are detached from its surface by using a piezoelectric actuator to vibrate the cover slip. Once a sphere released from the cover slip is trapped, the chamber is pumped to high vacuum and charge and recoil measurements begin. Due to the relatively short $T_{1/2}$ of $^{212}$Pb, no radioactive decay products will remain in the vacuum chamber after a few days following the loading of the cover slip.

\subsection{Electric charge measurement and control}
\label{sec:charge_control}
As shown in Fig.~\ref{fig:charge}, accurate measurement and control of the net electric charge of the sphere is important for detecting nuclear decays occurring within the sphere. The correlation between the motion of the sphere and an applied oscillating electric field is used to determine the net charge of the sphere. For this work, the oscillating electric field was applied in the $x$ direction, at a single frequency well above the resonance frequency of the sphere motion to allow the response to this field to be filtered out when reconstructing the momentum (see Sec.~\ref{sec:amp_recon}). The measured response is calibrated for each sphere by observing single $e$ steps induced by flashes of UV light or radioactive decays. For fixed drive amplitude and frequency, the calibration factor is stable within 10\% between the different spheres considered here. In addition to removing $e^-$ with the UV lamp, thermionic emission of $e^-$ from a tungsten filament is used to add $e^-$ to the sphere.

During data taking, an automatic procedure is used to maintain the magnitude of the net charge of the sphere at $|Q| \lesssim 50\ e$, which minimizes error in the charge reconstruction and avoids force noise arising from stray electric fields. Throughout data taking, the sphere's net charge is continuously monitored, and if the magnitude exceeds 50~$e$ either the tungsten filament or UV lamp is automatically activated by the data acquisition (DAQ) software. Since the nuclear decays within the sphere nearly always eject $e^-$ (see Fig.~\ref{fig:charge}), this automatic discharging occurs primarily from the filament. However, in rare cases the filament over charges the sphere in the opposite direction, and the UV lamp is activated to return the charge to $> -35\ e$. For each data file containing the measurements of the sphere motion, the DAQ program also records when the filament and UV lamp are activated.

\subsection{Impulse calibration}
\label{sec:calibration}
As described in the main text, the impulse response of each sphere is directly calibrated by applying a known electric impulse to the sphere in the $x$, $y$, and $z$ direction. This calibration provides an {\em in situ} characterization of the impulse response of the sphere. Since the net electric charge is exactly known (see Sec.~\ref{sec:charge_control}), accurate calibration of the applied impulse requires only knowledge of the electric field at the location of the sphere. 

\begin{figure*}
    \centering
    \includegraphics[width=0.8\textwidth]{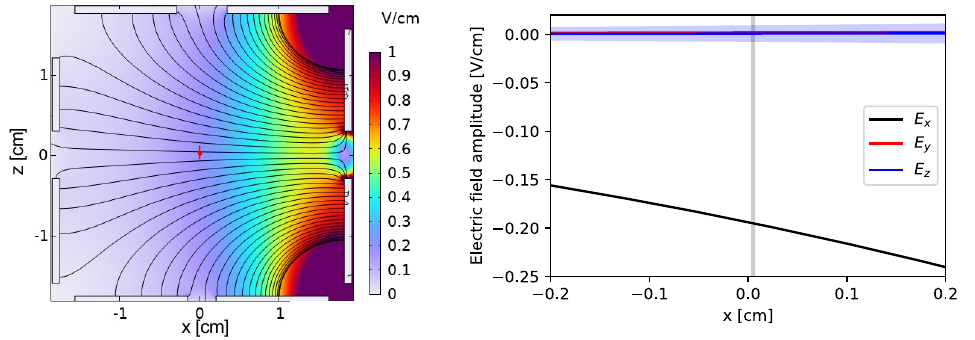}
    \caption{(left) Cross section in the $x-z$ plane of the 3D simulation of the electric field produced by a 1~V potential on the $+x$ electrode. The color scale shows the magnitude of the calculated electric field, while the streamlines indicate the field direction. The red point indicates the estimated sphere position and error (with $x$ error smaller than the marker size), as described in the text. (right) Simulated amplitude of the components of the electric field near the sphere location, for the potential configuration shown on the left. The uncertainty due to sphere position is indicated by the bands (comparable to the line thickness for the $E_x$ and $E_y$ components). The field primarily points in the negative $x$ direction, with components in the $y$ and $z$ directions at least an order of magnitude smaller. The vertical gray band indicates the measured sphere position and uncertainty in the $x$ direction.}
    \label{fig:efield_sim}
\end{figure*}

The electric field, as a function of applied voltage, is calculated using a finite element analysis (FEA) simulation of the electrode configuration with the COMSOL Multiphysics software program~\cite{comsol}. The exact geometry of the electrodes is modeled by the simulation, including features such as the holes in the electrode faces needed to allow the trapping and readout lasers, as well as the filament and UV light source for charge control. The electric field simulated for a $+$1~V potential applied to the $+x$ electrode with all other electrodes at ground is shown in Fig.~\ref{fig:efield_sim}. Using these simulations, the total impulse applied is determined after integrating the voltage waveform measured at the monitor output of the high voltage supply (Tabor 9400A). For each calibration, the sphere's motion in the $x$, $y$, and $z$ direction is measured while a waveform consisting of repeated square pulses with a pulse width of 1~ms and repetition rate of 2.3~Hz is applied. The voltage is set to provide a total impulse of approximately 200~MeV/c(comparable to the impulse during an $\alpha$ decay), i.e., typically a 6.7~V/cm electric field is applied during the pulse, with a net charge on the sphere set to be $|Q| = 1\ e$. A total of 300~s of data are recorded during each calibration, yielding approximately 700 pulses per calibration. The calibration procedure is repeated for pulses applied in the $x$, $y$, and $z$ directions individually. 

\begin{figure*}[b]
    \centering
    \includegraphics[width=\textwidth]{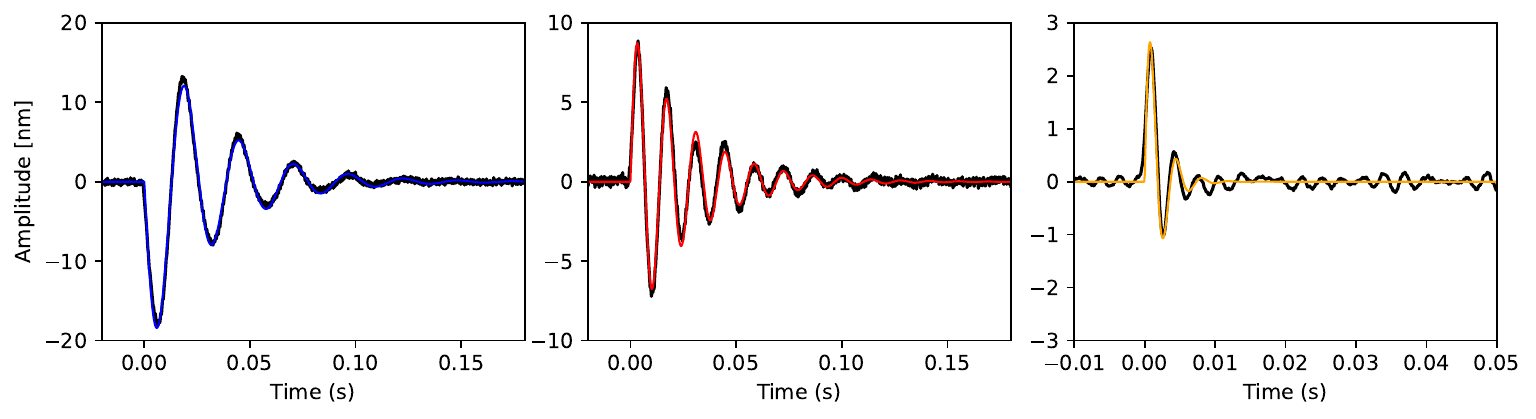}
    \caption{Example of the calibrated response for the $x$ (left), $y$ (center), and $z$ (right) directions. The black points show the averaged data from all calibration pulses, and the colored curves show the best fit to the impulse response described in the text. The best fit parameters from this calibration are summarized in Table~\ref{tab:res_params}.}
    \label{fig:impulse_resp}
\end{figure*}

Figure~\ref{fig:impulse_resp} shows the averaged impulse response from an example calibration, along with fits to a damped harmonic oscillator response: $s_i(t) = A\exp[-\gamma(t-t_0)]\sin[\omega_1 (t-t_0)]$ for time $t$ greater than the impulse time $t_0$, damping rate $\gamma$, and for $i = (x,y,z)$. For the underdamped case relevant here, $\omega_1 = \sqrt{\omega_0^2 - \gamma^2}$ for angular resonant frequency $\omega_0$. The best fit values of the resonator parameters are used to calibrate the waveforms from units of voltage recorded at the balanced photodiode to sphere position, assuming the impulse response function above for $A = I/(m \omega_1)$. Here $I$ is the applied impulse amplitude and $m$ is the sphere mass. 

Uncertainty in the sphere position relative to the electrodes can lead to uncertainty in the value of the applied impulse due to the non-uniformity of the electric field. Other systematic errors (e.g. uncertainties in the voltage pulse shape, uncertainties in the electrode geometry, and stray fields due to charge on dielectrics) are estimated to be sub-dominant to the sphere position uncertainty. The sphere position is measured relative to the electrode location by applying oscillating voltages to each pair of electrodes in the $\pm x$ and $\pm y$ directions and adjusting the relative amplitude of the voltage on each electrode to minimize the force on the sphere. Based on these calibrations, the sphere position is determined for the $x$ and $y$ directions to be $50 \pm 20\ \mu$m and $-40 \pm 20\ \mu$m respectively, with the origin at the center of the electrode cube. We expect the machining tolerances for the assembled face-to-face electrode separation to be approximately $\pm 150\ \mu$m, which is larger than the uncertainty estimated in the offsets above. However, this {\em in situ} calibration measures the sphere position relative to the as-assembled electrode center, so the impact of the machining errors on the knowledge of the electric field at the sphere position is subdominant. For the $z$ direction, the sphere position is measured to be $160 \pm 800\ \mu$m from a photograph of the trapped sphere location relative to the center hole on the electrodes.

Using these positions, we estimate the magnitude of the electric field at the sphere location and its corresponding uncertainty using the COMSOL simulation. The resulting electric fields are determined to be $E_x = -19.4 \pm 0.2$~V/m, $E_y = -20.9 \pm 0.2$~V/m, and $E_z = -20.2 \pm 2.1$~V/m, for a 1~V potential applied to the $+x$, $+y$, and $+z$ electrodes, respectively. The cross-talk (i.e., maximum magnitude of the electric field components perpendicular to the components above) is estimated from the simulation to be $<10$\% of the field magnitude in the applied direction, in all cases above.

Finally, there is uncertainty in the sphere mass since the radius and density are not precisely measured for each sphere, and previous studies have found lower sphere densities than expected for pure fused silica~\cite{PhysRevApplied.12.024037}. While an advantage of the calibration method described above is that this uncertainty does not affect the impulse calibration, it may affect the calibration of position that appears for reference in figures throughout this work. For the position calibration, we assume a nominal density of 2.0~g/cm$^3$~\cite{PhysRevA.96.063841} for the sphere and a radius of 1.4~$\mu$m, leading to an estimated sphere mass of 23~pg.  We estimate that there is a $\pm$30\% systematic uncertainty on the calibrated position response for a given sphere due to the uncertainty on its mass.

\begin{table}[t]
    \centering
    \begin{tabular}{c | c c}
        Coord. 	 & $f_0$ [Hz] & $\gamma/(2\pi)$ [Hz] \\
        \hline
        $x$ & 	$38.72\pm 0.02$ &	 $5.2\pm0.1$ \\
        $y$ & 	$72.41\pm 0.16$ &	 $5.9\pm0.2$ \\
        $z$ & 	$274.1\pm 1.6$ &	 $79\pm2.0$ \\
    \end{tabular}
    \caption{Examples of the best fit resonator parameters and errors from the calibration for the sphere shown in Fig.~\ref{fig:impulse_resp}. The $y$ resonant frequency is set to be higher than the $x$ resonant frequency using the applied feedback, which minimizes cross-talk between directions in the response of the optimal filter (see Sec.~\ref{sec:amp_recon}).}
    \label{tab:res_params}
\end{table}

\begin{figure*}[b]
    \centering
    \includegraphics[width=\textwidth]{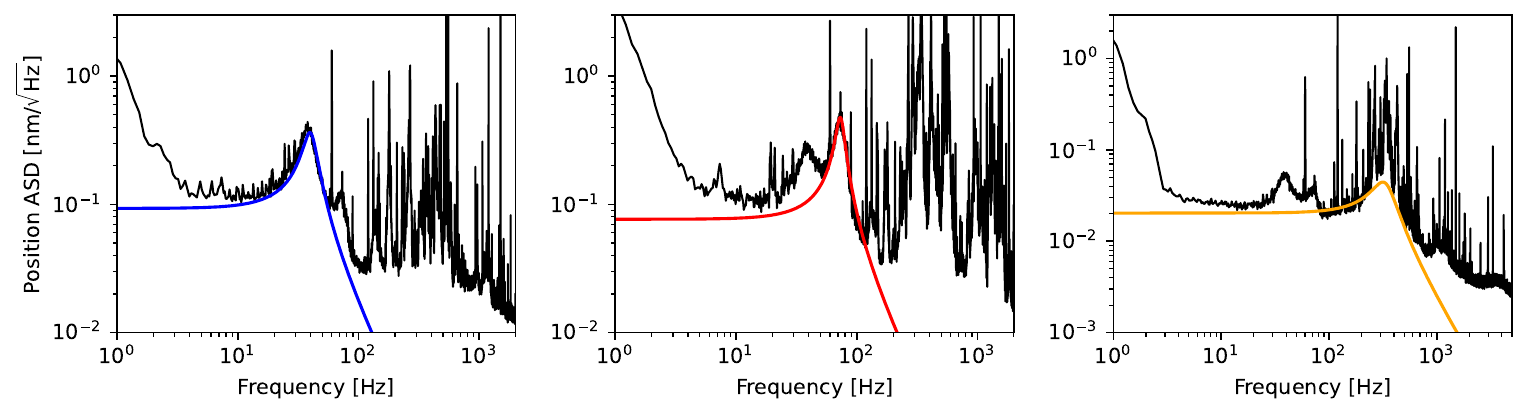}
    \caption{Example of position ASD for the $x$ (left), $y$ (center), and $z$ (right) directions measured by averaging 500~s of noise data. The colored lines correspond to the expected noise for a simple harmonic oscillator with the resonator parameters shown in Table~\ref{tab:res_params} driven by white noise. The displacement sensitivity of this system is $\lesssim 10^{-11}$~m/$\sqrt{\mathrm{Hz}}$.}
    \label{fig:noise}
\end{figure*}

\subsection{Noise}
\label{sec:noise}
Figure~\ref{fig:noise} shows examples of the amplitude spectral density (ASD), defined as the square root of the power spectral density (PSD), $J_i$, of the measured sphere position noise, for $i = (x,y,z)$. For the $x$ and $y$ directions, the noise near the resonant frequency is consistent with the response of a damped harmonic oscillator with parameters determined by the impulse calibration in Table~\ref{tab:res_params}. At frequencies between 100~Hz and several kHz, the noise is dominated by narrow lines, which are expected to arise from vibrational noise or electronic pickup. These noise lines substantially affect the sensitivity of the $z$ direction, degrading its sensitivity relative to the $x$ and $y$ directions. As described in Sec.~\ref{sec:amp_recon}, for the impulses considered in this work (which have a broad frequency spectrum) the data are analyzed taking into account these measured noise spectra and optimally fitting the pulse waveforms while weighting by the signal-to-noise at each frequency.

\begin{figure*}[t]
    \centering
    \includegraphics[width=\textwidth]{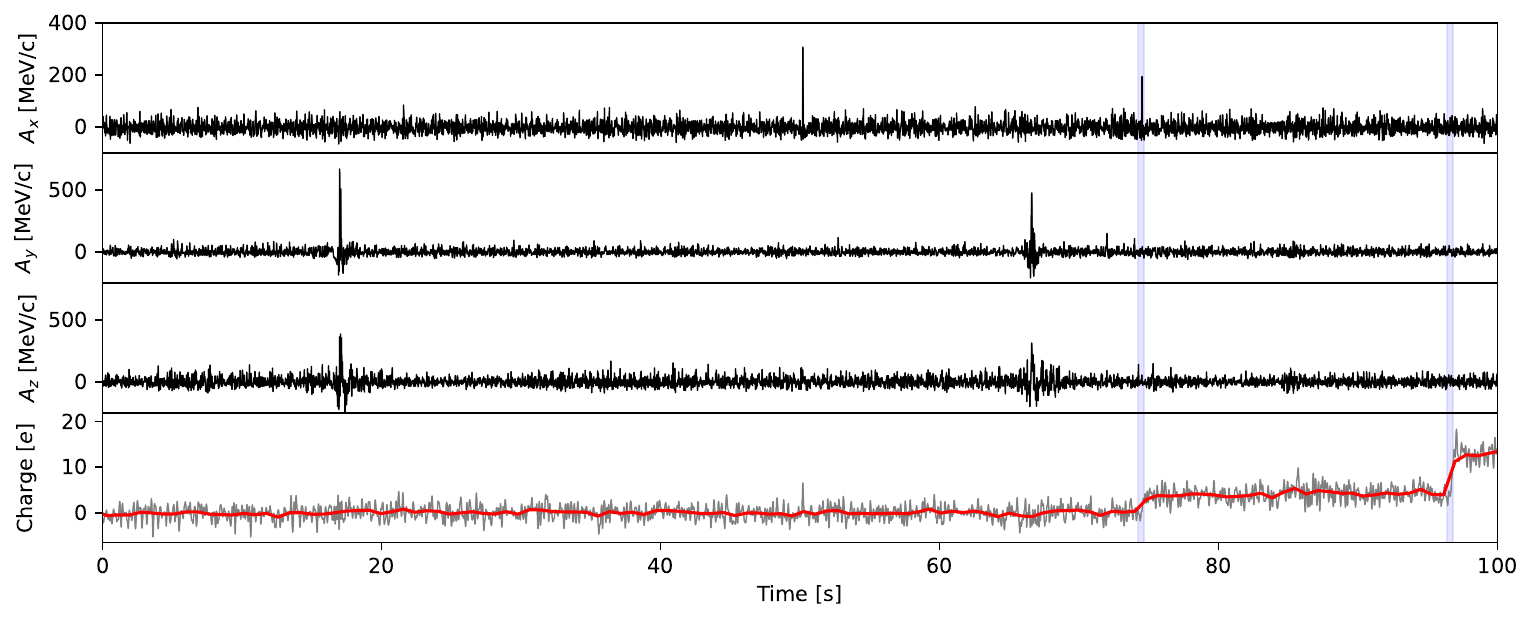}
    \caption{Example event containing candidate $^{212}$Bi--$^{208}$Tl coincident decays, in addition to transient noise that is not correlated with a charge change of the sphere. The upper three panels show the reconstructed impulse amplitude versus time in the $x$, $y$, and $z$, directions, while the lower panel shows the reconstructed charge. Charge changes observed around 75~s and 97~s are highlighted by the blue shaded regions.}
    \label{fig:transient}
\end{figure*}

In addition to the stationary noise whose ASD is shown in Fig.~\ref{fig:noise}, random pulses occur throughout the data taking period that are consistent with transient impulses or other force noise that is not correlated with a decay within the sphere. In past work, such impulses were seen to arise from vibrationally or acoustically transmitted noise sources that exert a force on the sensor~\cite{PhysRevLett.125.181102}. An example is shown in Fig.~\ref{fig:transient}. Here, in addition to the impulse seen around 75~s (a candidate $^{212}$Bi $\alpha$ decay), which is correlated with a sphere charge change, a large signal consistent with the expected pulse shape for an impulse in the $x$ direction is seen around 50~s, but with no charge change. In addition, noise bursts inconsistent with the impulse response are seen in the $y$ and $z$ channels around 18~s and 65~s. No impulse is seen at the time of the observed charge change around 97~s (consistent with the undetectable momentum of a candidate $\beta$ decay from $^{208}$Tl).

To characterize these transient noise sources, the same pulse reconstruction procedure applied in the search window around each observed charge change was instead applied at randomly selected times throughout the data taking period. This allows the probability of a random coincidence between these transient impulses and an observed charge change to be empirically characterized from the data itself. The gray shaded distribution in Fig.~\ref{fig:spectrum} shows the pulse reconstruction procedure applied to these random time windows. The non-Gaussian tail of this distribution characterizes the probability of random coincidences with large amplitude impulses. This distribution indicates there is low probability that the events in Fig.~\ref{fig:spectrum} above 150~MeV/c arise from random coincidences with transient noise.

\subsection{Momentum reconstruction}
\label{sec:amp_recon}
The calibration procedure described in Sec.~\ref{sec:calibration} provides an empirical method for directly characterizing the response of the sphere to the instantaneous impulses of interest here. However, since the frequency spectrum of such impulses is broad, and the noise spectrum measured in Fig.~\ref{fig:noise} contains low frequency noise and narrow vibrational lines above the damped harmonic oscillator response, several reconstruction methods were investigated to optimally separate signals from noise. 

\begin{figure*}
    \centering
    \includegraphics[width=0.8\textwidth]{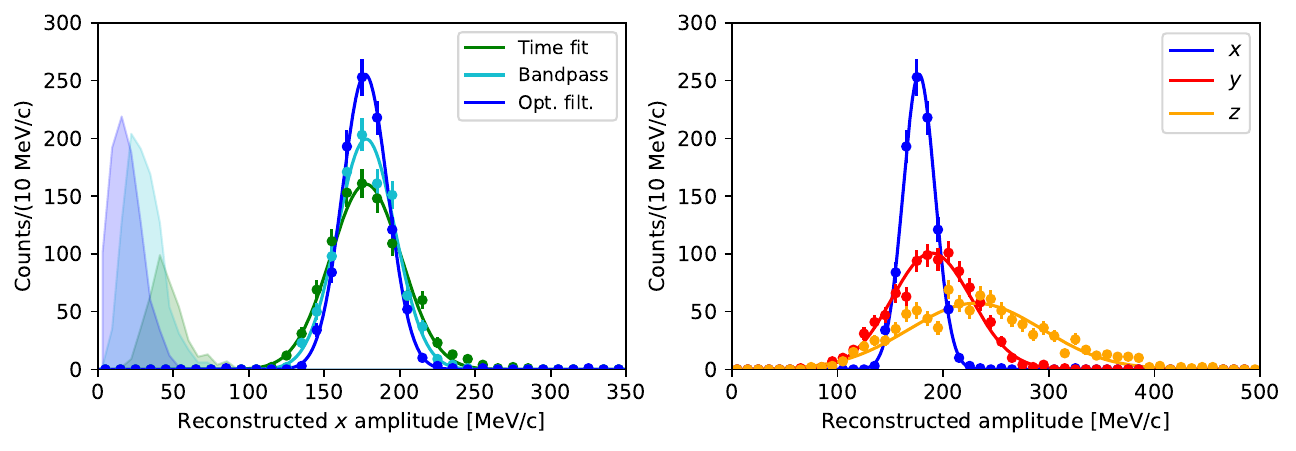}
    \caption{(left) Reconstructed amplitude distribution when applying different reconstruction algorithms to the impulse calibration data for an impulse of $I_x = 177 \pm 2$~MeV/c applied in the $x$ direction. The optimal filter provides the best resolution for the reconstructed amplitude with $\sigma_x = 15.0$~MeV/c. The bandpass filter ($\sigma_x = 18.5$~MeV/c) and time domain fit ($\sigma_x = 23.2$~MeV/c) perform similarly, although with slightly poorer resolution. The light shaded regions show the same algorithms applied to noise (rather than calibration impulses), indicating that the noise distribution is also optimally separated from the signals for the optimal filter. (right) Comparison of the reconstructed calibration amplitudes for impulses applied separately in the $x$, $y$, and $z$ directions using the optimal filter. The differences in impulse amplitude between directions arise from the different electric field at the sphere position (as described in Sec.~\ref{sec:calibration}). The $y$ and $z$ directions show poorer resolution than the $x$ direction ($\sigma_y = 37$~MeV/c and $\sigma_z = 62$~MeV/c, respectively).}
    \label{fig:recon_res}
\end{figure*}

Figure~\ref{fig:recon_res} compares several methods of pulse reconstruction for the sphere calibration shown in Figs.~\ref{fig:impulse_resp}--\ref{fig:noise}. Each method makes use of knowledge of the measured impulse response signal shapes, $s_i$, the noise power spectral densities, $J_i$, or both. Three methods are considered: 1) a time-domain fit of the waveform to the calibration templates $s(t)$; 2) the maximum pulse amplitude following application of a bandpass filter that is chosen to maintain the highest signal-to-noise region around the resonant frequency, while rejecting noise at lower and higher frequencies; and 3), an ``optimal filter'' that is constructed using the Fourier transform of the signal templates, $\tilde{s}_i$, and noise PSDs, $J_i$, to optimally weight the signal-to-noise in the frequency domain~\cite{Gatti:1986cw}. Specifically, the optimally filtered waveform is constructed for each recorded waveform, $x_n$ (with $n = 0,1,...,N-1$ indicating the time index of the sample) as:
\begin{equation}
\label{eq:of}
    A_{j,n} = \sum_{k=0}^{N-1} \frac{\tilde{s}_{j,k}^* \tilde{x}_{j,k} e^{2 \pi i k n}}{J_{j,k}}
\end{equation}
where the amplitude can be calculated separately for each coordinate $x_j$ for $(x_1, x_2, x_3) = (x,y,z)$. The expression is evaluated in the frequency domain where the noise at different frequencies is expected to be uncorrelated, and tildes indicate Fourier transforms. The template waveform is initially constructed to correspond to an impulse at $n=0$, such that the phase factor shifts the template waveform in time and allows the optimal filter to be calculated for an arbitrary impulse time. For computational efficiency, Eq.~\ref{eq:of} is implemented numerically via a fast Fourier transform (FFT) as the inverse FFT of $\tilde{\phi}\tilde{x}$ for the filter $\tilde{\phi} = \tilde{s}^*/J$. As shown in Fig.~\ref{fig:recon_res}, while all reconstruction algorithms considered are able to effectively reconstruct the $\alpha$ impulses of interest (with momenta between $\sim$200--250~MeV/c), the optimal filter performs the best in terms of both the reconstructed resolution and the mis-reconstruction of noise, and is used throughout this work.

\subsection{Correction for impulses induced by charge changes}
\label{sec:charge_kicks}
In addition to the impulse induced by the daughter nucleus stopping within the sphere, a small impulse is applied to the sphere at the decay time by the oscillating electric field used to monitor the net charge of the sphere (see Sec.~\ref{sec:charge_control}). If not accounted for, this unwanted impulse can provide a systematic error on the measurement of the recoil momentum. To minimize the impact of the field induced impulse, the frequency used to monitor the charge is typically chosen to be well above the resonant frequency of the sphere in $x$, and the minimum amplitude of this oscillating field sufficient to resolve the time of charge changes within $<100$~ms is chosen. For the data in this work, either a drive frequency of 111~Hz or 151~Hz was used depending on sphere, with an amplitude between 0.375--0.75~V (providing an electric field of 0.08--0.16 V/cm). While most of the force on the sphere due to this oscillating field occurs at the drive frequency, and is thus easily excluded from the impulse analysis, due to the step in the amplitude of this oscillating force at the decay time, a small broad-band impulse also occurs. 

\begin{figure*}
    \centering
    \includegraphics[width=\textwidth]{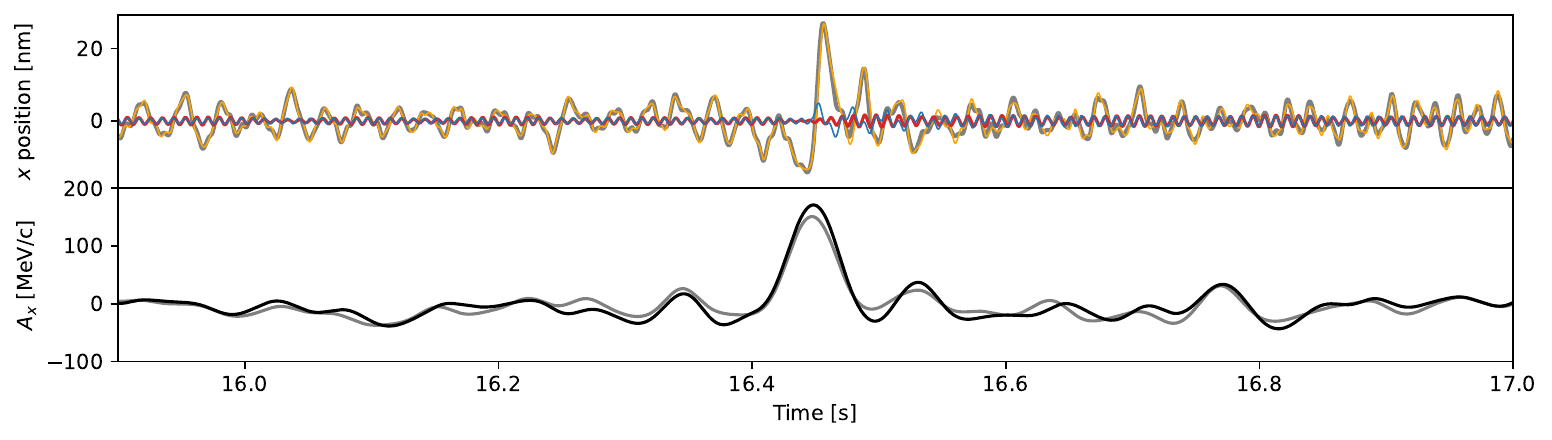}
    \caption{Example of the effect of the subtraction of the charge monitoring force for the example event in Fig.~\ref{fig:pulse}. (top) Measured position versus time before (gray) and after (orange) subtraction of the expected contribution to the motion from the oscillating electric field used to monitor the charge. Overlaid in red is the measured $x$ position after applying a bandpass filter $\pm$2~Hz around the driving frequency of 111~Hz for this sphere. A small increase in the amplitude of this motion and phase shift by $\pi$ is observed at the decay time, corresponding to the change in the electric charge from $-7\ e$ to $+13\ e$ at the decay time. The blue waveform shows the predicted response of the sphere to this drive, including the broad band impulse that occurs at the decay time. Subtraction of the blue curve from the position waveform gives the best estimate of the response to the nuclear recoil alone (orange). (bottom) Corresponding optimally filtered waveform before (gray) and after (black) subtraction of the blue curve from the position waveform. This subtraction has only a small impact on the reconstructed nuclear recoil amplitude.}
    \label{fig:kick_sub}
\end{figure*}

This impulse is in general a function of the charge of the sphere before and after the decay, and the phase of the oscillating electric field at the time of the decay. However, from the impulse response calibrated in Sec.~\ref{sec:calibration}, the decay time reconstructed from the recoil response (with $<10$~ms error), and the charge measured both before and after the decay, the unwanted impulse can be accurately subtracted from the waveform, before reconstruction of the impulse from the nuclear recoil using the methods described in Sec.~\ref{sec:amp_recon}. Figure~\ref{fig:kick_sub} shows an example of this subtraction procedure for the event plotted in Fig.~\ref{fig:pulse}. 

This correction can either increase or decrease the reconstructed amplitude, and is typically $<15$\% of the reconstructed amplitudes plotted in Fig.~\ref{fig:spectrum}. I.e., while the impulse provided by the oscillating electric field can be non-negligible, the effect is sub-dominant to the noise prior to correction and is expected to provide a negligible systematic error in the amplitude reconstruction of the nuclear recoil after correction. To verify the accuracy of this correction, a dedicated calibration was performed where a voltage waveform with a discrete step in the amplitude of the monitoring drive at 111~Hz was repeatedly applied, and the sphere response amplitude was reconstructed using the optimal filter. The subtraction procedure shown in Fig.~\ref{fig:kick_sub} was verified to reduce the reconstructed impulse by a factor of $> 10\times$. For a voltage step corresponding to a 20~$e$ charge change and the largest monitoring voltages considered here, the mean reconstructed impulse amplitude arising from the oscillating electric field after subtraction was found to be smaller than the reconstruction noise estimated in the calibration, indicating that we do not expect a significant systematic error in the reconstructed amplitude to arise from this driving field.

\subsection{Decay chain simulation}
\label{sec:decay_chain_MC}
As shown in Fig.~\ref{fig:implant_depth}, we expect the activity loaded within the sphere will be implanted within $\lesssim 60$~nm of the sphere surface. Lower chain $\alpha$ decays following the initial $^{212}$Pb implantation can thus in some cases eject the daughter nucleus from the surface of the sphere (which may additionally eject secondary O or Si nuclei). In addition, because of the poorer momentum resolution in the $y$ and $z$ directions (and uncertainty due to time-dependent drifts in the calibration for these directions), in this work, we focus on accurately reconstructing only the portion of the momentum for each decay the $x$ direction. A Monte Carlo simulation is used to both account for the escape of nuclear recoils throughout the decay chain and determine the expected projected momentum in $x$ for each simulated decay. 

\begin{figure*}
    \centering
    \includegraphics[width=\textwidth]{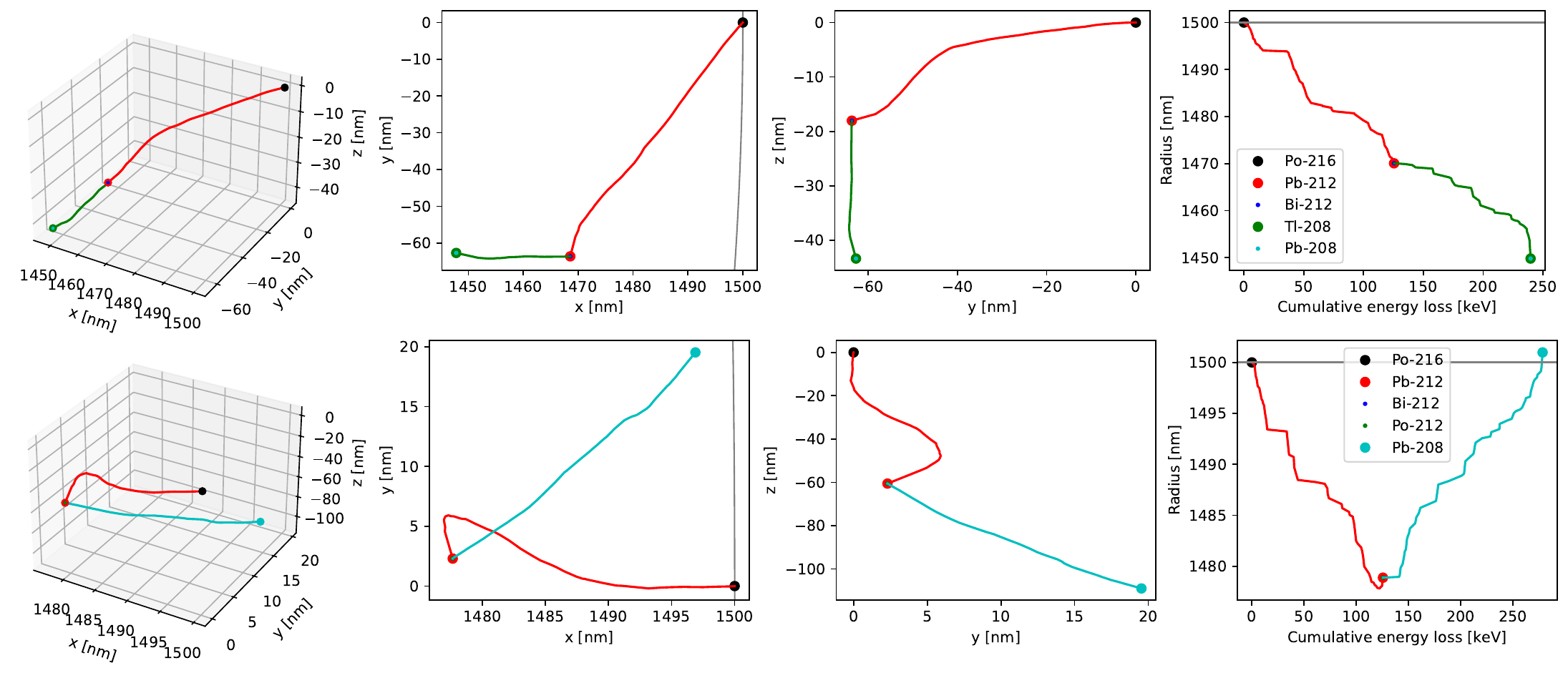}
    \caption{Example simulated decay events for the full decay chain starting from a randomly oriented $\alpha$ decay of $^{216}$Po at the sphere surface. The upper row of plots shows an example event with a $^{212}$Bi $\alpha$ decay, which reaches stable $^{208}$Pb without the nuclear recoil exiting the sphere. The lower plots show a decay chain with a $^{212}$Po $\alpha$ decay, for which the daughter $^{208}$Pb nuclear recoil exits the sphere. For each event, the left plot shows the 3D trajectory, with the projected paths in the $x$-$y$ and $y$-$z$ planes shown in the middle panels. The gray line indicates the edge of the sphere (with radius $1.5\ \mu$m). The right panel shows the radius of the daughter as a function of its energy loss throughout the trajectory. Daughters of $\alpha$ decays are shown as large markers. Nuclear recoils from $\beta$ decays (small markers) will not travel significant distance from the decay location and their trajectories are not simulated in detail.}
    \label{fig:example_events}
\end{figure*}

The momentum spectrum used to fit the observed data in Fig.~\ref{fig:spectrum}, was calculated using a simulation of the full decay chain based on SRIM. This simulation was used to produce $10^7$ decays starting from $^{216}$Po deposited at a random location on the surface of the sphere. This simulation makes use of the half-life, recoil energy, and branching ratio data from the NuDat 3.0 database~\cite{nudat}. For each $\alpha$ decay in the chain, a trajectory is first simulated with SRIM, and custom code is used to follow this trajectory through the geometry of the sphere until the nuclear recoil stops within the sphere or exits its surface. If the nuclear recoil exits the sphere, its remaining momentum upon exiting the sphere is recorded and the simulation terminates for that event. If the nuclear recoil stops, subsequent decays are then modeled until they either reach stable $^{208}$Pb at the end of the decay chain or the subsequent daughters exit the sphere. Examples of simulated events using this code are shown in Fig.~\ref{fig:example_events}. The loss of energy and momentum by the $\alpha$ within the sphere was also simulated in SRIM. For the worst case of transiting through the entire 3~$\mu$m thickness of the sphere, the energy loss within the sphere is $<350$~keV from the SRIM simulation, which corresponds to a maximum change in the momentum of the exiting  of $<$5~MeV/c. Since even this maximum change is small compared to the resolution, and the median decay (after averaging over all decay angles) will have a momentum loss $<$1 MeV/c, scattering within the sphere is neglected in the simulation. For simplicity the $\alpha$ is assumed to exit the sphere with momentum equal to its momentum at the decay location, with only the nuclear recoils of the decay daughters simulated in detail.

\begin{figure*}
    \centering
    \includegraphics[width=0.8\textwidth]{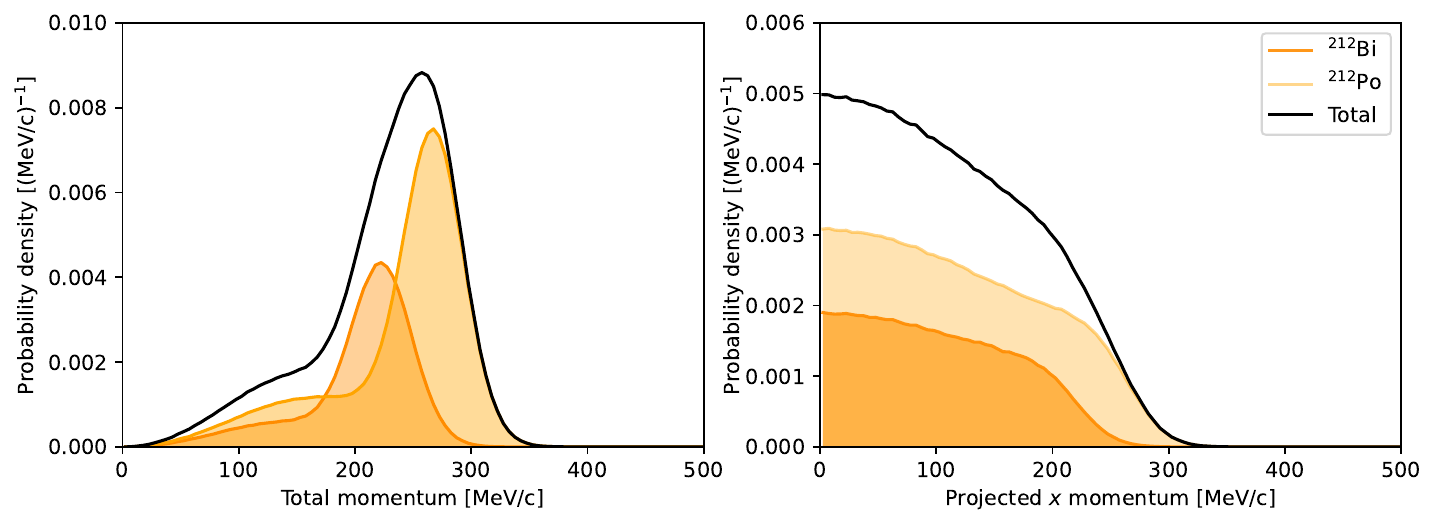}
    \caption{(left) Simulation of the magnitude of the momentum transferred to the sphere from the Monte Carlo described in the text, assuming an equal momentum resolution independent of the decay direction. The peak in the momentum distribution corresponds to $^{212}$Bi (dark orange) and $^{212}$Po (light orange) $\alpha$ decays for which the nuclear recoil stops in the sphere and its full momentum is transferred to the sphere's center-of-mass motion. Due to the shallow implantation of the $^{212}$Pb parent, approximately 25\% of decays have reduced momentum due to escape of the nuclear recoil from the sphere (providing the low momentum tail below the peak). (right) Projection of the $x$ component of the simulated momentum for each event. }
    \label{fig:momentum_mc}
\end{figure*}

The total momentum transferred to the sphere in each decay is calculated in the simulation using momentum conservation as $\vec{p}_{sph} = -(\vec{p}_\alpha + \vec{p}_{NR})$, where $\vec{p}_\alpha$ is the momentum of the $\alpha$ particle and $\vec{p}_{NR}$ is the momentum of the nuclear recoil simulated by SRIM for each decay upon exiting the sphere (if the nuclear recoil stops within the sphere, $\vec{p}_{NR} = 0$). Each component of $\vec{p}_{sph}$ is then smeared by a noise value randomly selected from the resolution measured in the {\em in situ} calibrations for each sphere, weighted by live time. The resulting momentum distribution is shown in Fig.~\ref{fig:momentum_mc} (left), for an example where we assume the noise in all three degrees of freedom is improved to the currently measured noise in the $x$ direction. The low momentum tail of the distribution corresponds to events in which the daughter nuclear recoil escapes before fully transferring its momentum to the sphere, due to the shallow implantation depth of the $^{212}$Pb in this work. Future work will aim to characterize the escape of these daughters, and possibly veto them based on the corresponding charge change. Alternative sphere synthesis methods could also allow the portion of the sphere containing the activity to be capped by a $\gtrsim$100~nm thick inactive silica layer, preventing escape of these daughter recoils~\cite{PRXQuantum.4.010315}. 

For each event, the $x$ projection of the momentum is also calculated, as shown in Fig.~\ref{fig:momentum_mc} (right). The low momentum tail of events in the projected distribution includes escaping daughter recoils as well as events in which the momentum transfer is not aligned in $x$. Future work to integrate conventional charged particle detectors around the trap to determine the recoil direction or improve the resolution in the $y$ and $z$ directions would allow these low momenta tails to be eliminated.

In addition to the $\alpha$ decays, from the simulation we expect $\sim$58$\%$ of all decays within the $^{212}$Pb decay chain to be $\beta$ decays (after accounting in the simulation for ejection of some $^{208}$Tl daughters of the $^{212}$Bi prior to their decay). In addition, 
the $^{212}$Po $\alpha$ decay is not resolvable from the preceding $^{212}$Bi $\beta$ decay due to its short half-life. 
With a typical recoil momentum of 1--3~MeV/c, the momentum carried by these decay products is significantly smaller than the best resolution of the momentum reconstruction algorithms used here (see Sec.~\ref{sec:amp_recon}), and the simulation neglects the momentum of any $\beta$, $\gamma$, or $\nu$ particles emitted in coincidence with the $^{212}$Po $\alpha$. 

In addition, for isolated $\beta$ decays (i.e. $^{212}$Pb or $^{208}$Tl), the change in the net charge of the microsphere will still be detected, and the optimal filter will be used to reconstruct the coincident impulse amplitude. This is expected to produce a distribution of reconstructed events with an amplitude consistent with noise. 
To accurately model the reconstructed amplitudes for these $\beta$ decays, we sample noise data uniformly distributed throughout the data acquisition period and apply the same optimal filter based reconstruction at randomly selected times (uncorrelated with a charge change) that we apply to detect recoils in coincidence with a charge change. This empirically driven procedure accounts for any effects from transient noise, random coincidence with noise bursts, etc. that may be present in the data (see Sec.~\ref{sec:noise}). 

The resulting distribution of reconstructed amplitudes for these random times is used to produce the gray shaded distribution in Fig.~\ref{fig:spectrum}. A total momentum distribution is then created by adding this to the results of the $\alpha$ particle recoil simulation. This distribution is then fit to the data as shown in Fig.~\ref{fig:spectrum}, with the ratios of the $^{212}$Po and $^{212}$Bi fixed based on the decay chain simulation, but allowing the rate of the modelled $\beta$ decays to vary as a free parameter in the fit.

\end{document}